\documentclass{JHEP3}
\usepackage{epsfig}

\usepackage{amsmath}

\title{Large entropy production inside black holes: a simple model}

\author{Gavin Polhemus \\
	JILA, Box 440, University of Colorado, Boulder CO 80309, U.S.A. \\
	E-mail: \email{gavin.polhemus@colorado.edu}
}
\author{Andrew J S Hamilton \\
	JILA, Box 440, University of Colorado, Boulder CO 80309, U.S.A.\\
	Dept.\ Astrophysical \& Planetary Sciences, Box 391, \\
	University of Colorado, Boulder CO 80309, U.S.A.\\ 
	E-mail: \email{Andrew.Hamilton@colorado.edu}
}

\newcommand{\col}{_{*}}
\newcommand{\abs}[1]{\left| #1 \right|}
\newcommand{\infinity}{\infty}
\newcommand{\textfrac}[2]{{\textstyle\frac{#1}{#2}}}

\newcommand{\sub}[1]{_{\textrm{#1}}}
\newcommand{\half}{\textfrac{1}{2}}

\newcommand{\units}[1]{{\renewcommand{\.}{\!\cdot\!}\rm\,#1}}
\newcommand\w{\omega}
\newcommand{\form}[1]{\boldsymbol{\mathsf #1}}

\abstract{
Particles dropped into a rotating black hole can collide near the inner horizon with enormous energies.  The entropy produced by these collisions can be several times larger than the increase in the horizon entropy due to the addition of the particles.
In this paper entropy is produced by releasing large numbers of neutrons near the outer horizon of a rotating black hole such that they collide near the inner horizon 
at energies similar to those achieved at the Relativistic Heavy Ion Collider.  The increase in horizon entropy is approximately 80 per dropped neutron pair, while the entropy produced in the collisions is 160 per neutron pair.
The collision entropy is produced inside the horizon, so this excess entropy production does not violate Bousso's bound limiting the entropy that can go through the black hole's horizon.  The generalized laws of black hole thermodynamics are obeyed.
No individual observer inside the black hole sees a violation
of the second law of thermodynamics
}

\keywords{Black Holes}

\preprint{}

\begin{document}

\section{Introduction}\label{introduction}

Adding particles to a black hole invariably increases its area,
and therefore its entropy.
Dissipation inside the horizon, for example by partlcle collisions,
can produce more entropy, even many times more entropy \cite{Wallace:2008zz},
than the increase in the Bekenstein-Hawking entropy of the black hole.
This might appear to be a cause for concern, but if fact the excess
entropy is consistent with black hole thermodynamics and statistical mechanics
\cite{Polhemus:2009vv}, and does not violate covariant bounds on the entropy that can pass through a light sheet \cite{Bousso:2002ju} (as should be expected given the theorem of \cite{Flanagan:1999jp}).
 
The purpose of this paper is to investigate a model of entropy
production inside black holes that involves experimentally established
particle physics, as observed at the Relativistic Heavy Ion Collider
(RHIC) \cite{Back:2004je, Adler:2004zn}.
We drop particles into a rotating, macroscopic Kerr black hole,
such as astronomers observe.
The specific angular momentum should be large, but less than extremal,
$a \leq M$.
Numerical calculations below will use a = 4/5 M.

We drop particles into the black hole in the spirit of models
considered by Bekenstein \cite{Bekenstein:1973ur}, Bousso \cite{Bousso:2002ju}, Wald \cite{Wald:1999vt}, and others.
A single-layer belt of closely packed neutrons is lowered onto the equatorial region of the black hole's outer horizon and dropped.
The belt is lowered until all of the neutrons are approximately one neutron radius away from the horizon, a proper distance of $1.2\units{fm}$.
Before it is dropped, the belt is given angular momentum in the same direction as the black hole's spin, so that the neutrons become outgoing before crossing the inner horizon.  A few black hole crossing times later another identical belt is dropped in the same manner, but with opposite angular momentum.  The second belt will remain ingoing and crash into the earlier belt near the inner horizon. The motion of the falling neutrons is reviewed in section \ref{review}.  

The energy of the collisions grows exponentially with the delay between the drops.  Waiting a few black hole crossing times before releasing the second belt results in collisions near the inner horizon which are comparable to nuclear collisions at the RHIC.  The entropy produced by these collisions is approximately double the increase in the horizon entropy.  
The horizon's entropy increase is calculated in section \ref{horizon}.  The entropy produced in the collisions is calculated in section \ref{collisions}.

The large entropy produced near the inner horizon cannot be seen by observers outside the black hole, so the laws of black hole thermodynamics are not affected for outside observers.
While the laws of thermodynamics are obeyed for all observers, the excess entropy does have consequences for understanding the black hole's microscopic degrees of freedom.  These issues are discussed in \cite{Polhemus:2009vv}. 

We work in Planck units, $k_B = c = G = \hbar = 1$.

\section{Quick review of equatorial orbits about a rotating black hole}\label{review}

The Boyer-Lindquist line element for a rotating black hole of mass $M$ and specific angular momentum $a$ is
\begin{equation}
	ds^2 = -\frac{\Delta}{\rho^2}\left[dt - a \sin^{2}\theta \,d\phi\right]^{2}
	+ \frac{\rho^2}{\Delta}\,dr^2
	+ \rho^2\, d\theta^2
	+ \frac{\sin^2\theta}{\rho^{2}}\left[(r^{2}+a^{2})\,d\phi - a\,dt\right]^2\,,
\end{equation}
where
\begin{equation}
	\Delta \equiv r^2 - 2Mr + a^2
	\qquad\text{and}\qquad
	\rho^2 \equiv r^2 + a^2\cos^2\theta\,. 
\end{equation}
We will be considering a black hole which has $a = \textfrac{4}{5}M$ as an example.

The two zeros of the horizon function, $\Delta$, are the inner and outer horizons, located at $r_{\pm} = M \pm \sqrt{M^{2} - a^{2}}$.
The horizon function is positive outside the outer horizon, and negative inside.  ($\Delta$ is formally positive again inside the inner horizon, but instabilities cause the inner horizon to become singular, truncating the space-time at the inner horizon.)
The line element can be written in terms of an orthonormal tetrad of basis 1-forms,
\begin{equation}
	ds^2 = \mp (\form\w^t)^2 \pm (\form\w^r)^2 + (\form\w^\theta)^2 + (\form\w^\phi)^2\,.
\end{equation}
The top signs apply outside the outer horizon, where the $t$-coordinate is time-like and the $r$-coordinate is space-like.  The lower signs apply inside the outer horizon, where the $t$-coordinate is space-like and the $r$-coordinate is time-like.

The tetrad basis 1-forms can be read off of the coordinate line element.
The inverse formulae will be used frequently (except for the $\theta$ component, which does not play a role in equatorial motion):
\begin{subequations}
\begin{align}
	\form dr &= \frac{\sqrt{\abs{\Delta}}}{\rho}\form\w^r\,,	\\
	\form d\phi &= \frac{1}{\rho}\left(\csc\theta\,\form\w^\phi
			+ \frac{a}{\sqrt{\abs{\Delta}}}\form\w^t\right)\,, \\
	\form dt &= \frac{1}{\rho}\left(\frac{r^2+a^2}{\sqrt{\abs{\Delta}}}\form\w^t
			+ a\sin\theta\,\form\w^\phi\right)\,.
\end{align}
\end{subequations}

The particle's motion is governed by four constants, its mass ($m$), energy-at-infinity ($E$), azimuthal angular momentum ($L_{z}$), and Carter integral ($\mathcal{Q}$).  The particles will be dropped in nearly equatorial orbits, where $\theta \approx \pi/2$.  The Carter integral is zero for equatorial orbits, so it will not appear until equation (\ref{eq.Carter}), where it is used to show that the falling neutrons will not stray far from the equatorial plane during their descent.

The particle's momentum is easily written in the coordinate basis using the conserved energy-at-infinity, $E$, and the conserved $z$-component of the angular momentum, $L_{z}$.  The radial potential, $\mathcal{R}$, parameterizes the $r$-component of the momentum.
For neutrons moving in the equatorial plane the momentum 1-form $\form{p}$ is
\begin{equation}
	\form p 
		= -E\,\form dt + L_{z}\,\form d\phi - \frac{\sqrt{\mathcal{R}}}{\abs{\Delta}}\,\form dr\,. \label{eq.p1}
\end{equation}
The $r$-component of the momentum is negative for dropped neutrons, since they move inexorably towards smaller radius.
In the tetrad basis the momentum becomes
\begin{equation}
	\form p 
		= \frac{1}{r}\left(\frac{-P}{\sqrt{\abs{\Delta}}}
			\form\w^t
			+ J\,\form\w^\phi
			+\frac{-\sqrt{\mathcal{R}}}{\sqrt{\abs{\Delta}}}\,\form \w^{r} \right), \label{eq.p2}
\end{equation}
where we have introduced
\begin{equation}
	J \equiv L_{z} - Ea
	\qquad\text{and}\qquad
	P \equiv Er^{2} - aJ\,.
	\label{eq.Phi}
\end{equation}
Outside the outer horizon, $P$ parameterizes the energy in the tetrad frame, while inside the outer horizon it parameterizes the radial momentum. 
The role of $\mathcal{R}$ is complementary, parameterizing the tetrad frame radial momentum outside the outer horizon, and the tetrad frame energy inside the horizon.  Since the tetrad frame energy is necessarily positive, $P$ is necessarily positive outside the outer horizon.

The radial momentum and the energy in the tetrad frame both diverge near the horizons due to the $\sqrt{\abs{\Delta}}$ in the denominator of equation (\ref{eq.p2}).  This signifies the particles accelerating towards the speed of light in the tetrad frame as they approach the horizon.

Inside the outer horizon, the sign of $P$ determines the direction of the particle's motion in the $t$-coordinate -- positive $P$ for ingoing neutrons which move in the negative $t$-direction, and negative $P$ for outgoing neutrons which move in the positive $t$-direction.  At the outer horizon $P$ is positive, so the neutrons are, unsurprisingly, traveling inwards.  However, $P$ can change sign during the fall from the outer horizon toward the inner horizon, causing the neutrons to approach the  inner horizon in the outgoing direction.  The huge entropy is produced when these outgoing neutrons, accelerating towards the speed of light in the outgoing direction, collide with ingoing neutrons, which are accelerating towards the speed of light in the ingoing direction.

$J$ parameterizes the particle's momentum in the $\phi$ direction as measured in the Boyer-Linquist tetrad basis.
In order for a neutron to change its radial motion from ingoing ($P>0$) to outgoing ($P<0$) $J$ must be positive.
The small change in the black hole's specific angular momentum due to the addition of a neutron is also proportional to $J$,
\begin{equation}
	da = \frac{J}{M}\,. \label{eq.da}
\end{equation}
This will be useful in computing the change the entropy of the horizon.

The radial potential $\mathcal{R}$ can be calculated using the fact that the magnitude of the particle's momentum is its mass.  This is easiest using the tetrad form of the momentum, equation (\ref{eq.p2}),
\begin{equation}
		\mathcal{R} = P^{2} - \left( m^{2}r^{2} + J^{2} \right)\Delta \label{eq.R}\,.
\end{equation}
Equations (\ref{eq.Phi}) and (\ref{eq.R}) relate all of the equatorial components of the motion to the conserved quantities and the radial coordinate, $r$.

\section{Dropping through the outer horizon: Horizon entropy}\label{horizon}
The neutrons are dropped from a proper height $b$ above the horizon.  The horizon function at the drop point, $\Delta_{b}$, is given by the near-horizon approximation,
\begin{equation}
	\sqrt{\Delta_{b}} \approx b \frac{\sqrt{M^{2} - a^{2}}}{r_{+}} \approx \textfrac{3}{8}b\,.
\end{equation}
At the moment the neutron is dropped it has no radial momentum, so $\mathcal{R} = 0$.  
The drop height determines the energy-at-infinity, $E$, by equations (\ref{eq.R}) and (\ref{eq.Phi}),
\begin{equation}
	E \approx \frac{3}{8}\frac{b}{r_{+}^{2}}\sqrt{m^{2}r_{+}^{2} + J^{2}} + \frac{a}{r_{+}^{2}}J\,.\label{eq.E1}
\end{equation}
The first term in equation (\ref{eq.E1}) is the energy due to the mass and azimuthal kinetic energy in the tetrad frame, red-shifted down by a large factor.  The second, much larger term in equation (\ref{eq.E1}) is due to the centrifugal potential.

The change in the black hole's mass due to the addition of the neutron is $dM = E$.  Since neutrons with both positive and negative $J$ will be added, the smaller first term in equation (\ref{eq.E1}) will be needed to calculate the change in horizon entropy.

Two bands with opposite $J$ will be dropped to produce collisions near the inner horizon.
The addition of these neutrons will cause a small increase in the black hole's Bekenstein-Hawking entropy and a larger collision entropy.
The increase in horizon area due to particles dropped near the horizon is fully worked out in \cite{Bekenstein:1973ur}.  The results relevant to present situation are quickly reproduced here.

Every neutron dropped with positive $J$ must be followed by a partner with negative $J$.  The resulting changes in the specific angular momentum, equation (\ref{eq.da}), exactly cancel,
\begin{equation}
	da =\frac{J_{1} + J_{2}}{M} = 0\,.
\end{equation}
The changes in the black hole's mass, $dM$, also largely cancel, so the subleading term in equation (\ref{eq.E1}) must be kept.  For the pair of neutrons,
\begin{align}
	dM & = E_{1} + E_{2}	\notag \\
		&\approx \frac{3}{4}\frac{b}{r_{+}^{2}}\sqrt{m^{2}r_{+}^{2} + J^{2}}\,.
\end{align}
The entropy of the horizon, $S\sub{BH} = A/4 = \pi (r_{+}^{2} + a^{2})$, changes by an amount
\begin{equation}
	dS\sub{BH} 
		\approx 4 \pi b \sqrt{m^{2} + \frac{J^{2}}{r_{+}^{2}}}\,.\label{eq.dS}
\end{equation}
The entropy increase can be made small by dropping the neutrons from a short proper distance  $b$ above the outer horizon.  However, this distance cannot be smaller than the size of the neutron, approximately $1.2\units{fm}$.
The mass of a neutron is $.94\units{GeV}$, leading to a minimum change in the horizon entropy $dS\sub{BH} \approx 72$ (using $1\units{GeV} = 5.05\units{fm^{-1}}$).  The minimum is only achieved for small $J$, but large $J$ will result in higher energy, and therefore higher entropy, collisions near the inner horizon.   A compromise to achieve low horizon entropy and high collision entropy is $J=\half mr_{+}$.  This gives an entropy increase on the horizon per neutron pair of
\begin{equation}
	dS\sub{BH}
	\approx 80 \label{eq.Sbh}\,.
\end{equation}
This entropy increase is independent of the mass and angular momentum of the black hole, and independent of angular momentum of the dropped neutrons when $J \ll mr_{+}$.  If the neutrons could be dropped from height equal to their Compton wavelength, then the entropy increase would be of order 1 in plank units \cite{Bekenstein:1973ur}.  However, the large size the neutrons requires a higher drop, leading to the larger minimum entropy in equation (\ref{eq.Sbh}).

\section{Collisions near the inner horizon: Collision entropy}\label{collisions}

At the Relativistic Heavy Ion Collider, the entropy created per colliding nucleon pair is roughly double the entropy calculated in equation (\ref{eq.Sbh}), as we will explain below.  In this section we first discuss the entropy created in RHIC collisions, and then show that the same entropy can be produced in RHIC-like collisions inside a rotating black hole.

Heavy ion collisions at RHIC create a quark-gluon plasma fireball.  The fireball is boost invariant along the collision axis over a large rapidity range, roughly spanning the range between the rapidities of the two colliding nuclei.  Due to the strength of the interactions, the thermalization time is approximately $\tau\sub{therm} = 1\units{fm}$ in the local rest frame of the quark-gluon plasma.  At thermalization, the entropy per rapidity in central collisions is approximately $dS/dy \approx 4500$ at mid-rapidity for RHIC collisions with nucleon-nucleon collision energies of $E_{*} = 200\units{GeV}$ \cite{Fries:2009wh}.  (A subscript ``${}_{*}$'' denotes collision values.)

Multiplying the entropy per rapidity by the width of the rapidity range for these collisions gives an total entropy per collision of $48\,000$.  The number of participating nucleons in a central Au+Au collision is about 350, so the entropy per colliding pair of nucleons is $S_{*} \approx 270$.  This is probably an overestimate of the entropy because the entropy per rapidity is lower at high rapidity.

A better estimate can be achieved by comparing the entropy to the number of observed charged particles \cite{Gubser:2008pc}.  Entropy in the thermalized plasma is a measure of the number of degrees of freedom, which is roughly proportional to the number of particles.  The total number of charged particles as well as the number of charged particles per pseudorapidity are well measured experimentally, and can be used to get a better estimate of the total entropy per nucleon pair.

In central Au+Au collisions at $E_{*} = 200\units{GeV}$ the number of charged particles per pseudo-rapidity is approximately $dN\sub{ch}/d\eta = 660$ \cite{Back:2004je}.  At mid rapidity the ratio of pseudorapidity to rapidity is approximately $d\eta/dy = 1.25$ \cite{Adler:2004zn}, leading to $dN\sub{ch}/dy = 820$.

Comparing the entropy per rapidity to the number of charged particles per rapidity gives an estimate ratio of total entropy to the total number of charged particles, 
\begin{equation}
	\frac{S_{*}}{N\sub{ch}} \approx \frac{dS/dy}{dN\sub{ch}/dy} \approx 5.5 \,.
\end{equation}
The number of charged particles per colliding nucleon pair is $N\sub{ch} \approx 29$ \cite{Back:2004je}.  This gives an estimate of the entropy created per nucleon pair,
\begin{equation}
	S_{*} \approx 160 \label{eq.Scol} \,.
\end{equation}
This entropy is roughly double the entropy that the pair of nucleons would create when dropped through the outer horizon, equation (\ref{eq.Sbh}).

It is expected that higher energy collisions will create even more entropy.  The  number of charged particles per pseudorapidity per colliding nucleon pair grows logarithmically with collision entropy \cite{Adler:2004zn},
\begin{equation}
	\frac{dN\sub{ch}}{d\eta} \approx .74 \ln\left(\frac{E_{*}}{m}\right) \,.
\end{equation}
The rapidity range also grows like $\ln(E_{*}/m)$ so that the entropy is 
\begin{equation}
	S_{*} \approx 5.7 \ln^{2}\left(\frac{E_{*}}{m}\right).
\end{equation}
Pb+Pb collisions at the Large Hadron Collider will have about 60\% higher rapidity than RHIC, resulting in an entropy production per nucleon about 2.6 times larger \cite{Adler:2004zn}, which is about $S_{*} = 410$, over five times more entropy than was created when the same nucleons fell through the outer horizon.  If the trend continues, the entropy released in GUT scale collisions would produce an entropy of $S_{*} \approx 7000$, about 90 times the entropy created on the horizon.

In order to create the large collision entropy inside a black hole, three things need to occur.  First, the neutrons have to be aimed so that they actually collide.  Second, the collision energy must be high enough.  Finally, the resulting fireball must have time to thermalize.

Dropping a huge number of neutrons ensures collisions.  Each neutron must be dropped near the horizon, so only one layer of neutrons can be dropped.  The neutrons do not need to lie exactly on the equator of the black hole, but can extend above and below the equator by several degrees.
Equatorial orbits are stable in the $\theta$ direction, so variation in the initial $\theta$ position and momentum in the $\theta$ direction will not grow as the neutrons fall.  This stability can be seen by looking at the fourth constant of the motion, $\mathcal{Q}$:
\begin{align}
	\mathcal{Q} &= p_{\theta}^{2} +
			\cos^{2}\theta\,\left[a^{2}(m^{2} - E^{2})
			 + \frac{L_{z}^{2}}{\sin^{2}\!\theta}\right] \notag \\
		&\approx p_{\theta}^{2}
			+ \left( \theta - \frac{\pi}{2} \right)^{2}
				\left[a^{2}m^{2} + \frac{3}{2}J^{2} \right]\,. \label{eq.Carter}
\end{align}
$\mathcal{Q}$ is quadratic in small deviations from $\theta = \pi/2$ or small momentum in the $\theta$ direction, so neutrons cannot spread in the $\theta$ direction absent other forces.  As the radius gets smaller, the layer of neutrons will get thicker.  
The strong nuclear force is likely to increase the effectiveness by further concentrating the neutrons in the vicinity of the equator.
Assuming their extent in the $\theta$ direction remains constant during the fall, the neutrons will create a layer approximately sixteen neutrons thick by the time they reach the inner horizon, about four times the thickness of a gold nucleus.  Few if any neutrons will pass through such a layer without colliding violently.

The neutron density will be decreased somewhat by decays.  The resulting protons will repel each other and leave the neutron band.  For decays to seriously decrease the neutron concentration, the neutrons would have to fall for several minutes, which will occur only in the largest astrophysical black holes.  Smaller holes provide sufficient energy, as we will now show.

Collisions close to the inner horizon ensure sufficient collision energy.
The neutrons colliding near the inner horizon will have equal and opposite angular momenta J relative to the tetrad frame.  As a result, the collision center of mass is at rest in the tetrad frame.  The collision energy, $E_{*}$, is the total tetrad frame energy of the colliding neutrons,
\begin{equation}
	E\col = \frac{2}{r_{-}}\frac{\sqrt{\mathcal{R}}}{\sqrt{-\Delta_{*}}}\,.
\end{equation}
Near the inner horizon the nucleus is relativistic in the tetrad frame, so $\sqrt{\mathcal{R}} = \abs{P} = \abs{Er_{-}^{2 }-aJ}$.
After the collision, all of the debris will fall to the inner horizon, where violent instabilities destroy everything \cite{Poisson:1990eh, Hamilton:2008zz}.
In the tetrad rest frame the proper time from the collision to the inner horizon, $\tau_{*}$, is given by a near horizon approximation
\begin{equation}
	\sqrt{-\Delta_{*}} \approx \tau_{*} \frac{\sqrt{M^{2} - a^{2}}}{r_{-}}
		\approx \textfrac{3}{2}\tau_{*} \,.
\end{equation}
This gives the collision energy in terms of the tetrad angular momentum $J$ and the proper time $\tau_{*}$,
\begin{equation}
	E\col \approx \frac{5}{2}\frac{J}{\tau_{*}}\,.
\end{equation}
To stay near the minimum horizon entropy, we have imposed $J\leq \half mr_{+}$ in equation (\ref{eq.Sbh}), which translates into a maximum time for the fireball to equilibrate, $\tau_{*} \leq \frac{5}{4} m r_{+}/E_{*}$.  Since the outer radius of the black hole is macroscopic, this is not an incredibly small time.  To achieve RHIC energies, $\tau_{*} \leq r_{+}/160$.  For GUT energy collisions, $\tau_{*} < r_{+}/10^{15}$.

The fireball must have enough time to thermalize before hitting these instabilities so that the entropy calculated in equation (\ref{eq.Scol}) actually occurs.  For QCD a time $\tau\sub{therm} \approx 1\units{fm}$ is sufficient.  However, some parts of the fireball are moving with rapidity comparable to the rapidity of the colliding neutrons.  As a result, the time for thermalization must be increased by a factor of $E_{*}/m$ so that all parts of the fireball can thermalize.
To have sufficient energy to create the entropy and sufficient time for thermalization, the proper time $\tau_{*}$ between the collision and the arrival of the inner horizon in the tetrad frame must satisfy
\begin{equation}
	\frac{E_{*}}{m} \tau\sub{therm} \leq \tau_{*} \leq \frac{5}{4}\frac{m}{E_{*}}r_{+}\,.
\end{equation}
Both of these equalities can be satisfied if the black hole is sufficiently large, 
\begin{equation}
	r_{+} \geq \frac25\left(\frac{E_{*}}{m}\right)^{2}\tau\sub{therm}\,.
\end{equation}
To achieve RHIC like collisions, where $E_{*}/m \approx 200$, even a microscopic black hole is large enough (although microscopic black holes may be too unstable).  GUT collisions, however, require an absurdly large black hole, with an outer radius of $10^{15}\units{m}$, or about one light month.  A supermassive black hole like the one at the center of the Milky Way could provide enough energy and thermalization time for collisions with energies of $10^{12}\units{GeV}$ per neutron pair, which would create roughly 50 times more entropy than is produced on the horizon.

The radius at which the collisions occur depends on the time delay between drops.  The longer the delay, the deeper the collision.  The change in the $t$-coordinate as the neutrons fall is given by 
\begin{equation}
	\frac{dt}{dr} = -\frac{r^{2}}{\Delta}\frac{P}{\sqrt{\mathcal{R}}}\,. \label{eq.dtdr}
\end{equation}
As the neutrons approach the inner horizon $dt/dr$ diverges.  Neutrons with positive $J$ go outward to $t=\infinity$ while neutrons with negative $J$ go inward to $t=-\infinity$.  Plugging in $P$ from equation (\ref{eq.Phi}) and $E$ from equation (\ref{eq.E1}) (ignoring the smaller first term) gives
\begin{equation}
	\frac{dt}{dr}
		= \frac{\mp r^{\frac32}(r+r_{+})}{
			(r-r_{-})(r_{+}-r)^{\frac12}
			\sqrt{
				\left( \frac{r_{+}}{r_{-}}\mu^{2} - 1 \right) r^{2}
				- \left( \mu^{2} + 1 \right) r_{+} r
				+ \left( \frac{r_{+}}{r_{-}} + 1 \right) r_{+}^{2}
			}
		}\,, \label{eq.dtdr2}
\end{equation}
where $\mu \equiv mr_{+}/J$ and the positive sign is for negative $J$.

The total change in the neutron's $t$-coordinate, from the drop point to the collision, is found by integrating equation (\ref{eq.dtdr2}) from the drop point to the collision radius.  Since the calculations for the ingoing and outgoing shells are identical but with opposite signs, the total change in their relative $t$-coordinate is double the integral of equation (\ref{eq.dtdr2}).  The particles must be at the same $t$-coordinate when they collide, so difference in times of the drops, $t_{12}$, must be equal to the total change in relative $t$-coordinate.
Equation (\ref{eq.dtdr2}) diverges at both horizons, but the integral diverges only at the inner horizon.  The integral can be separated into a sum of a logarithmically divergent part and a finite part:
\begin{equation}
	t_{12} 
		= \frac{2r_{-}^{2}}{r_{+}-r_{-}}\int_{r_{*}}^{r_{+}}\frac{dr}{(r-r_{-})} + Cr_{+}\, ,
\end{equation}
where numerical integration can be used to find $C$.  Using the specific angular momentum $a = \frac45 M$ and $J = \half mr_{+}$ gives
\begin{equation}
	t_{12} = r_{+}\left( \frac{1}{3}\ln\frac{2E_{*}}{5m} + 2.2 \right)	\,.
\end{equation}
The required delay time grows only logarithmically with desired collision energy, so very high energies can be achieved by waiting only a few crossing times.  RHIC energies, for example, can be achieved by waiting approximately 4 black hole crossing times.  GUT energies can be achieved by waiting 14 crossing times.

\section{Conclusions}
\label{conclusion}

We have shown that the entropy produced by particle collisions inside a black hole can exceed the 
Bekenstein-Hawking entropy produced by dropping those particles through the horizon.  The excess collision entropy cannot be observed from outside the black hole and has no impact on black hole thermodynamics as seen by outside observers.  The second law of thermodynamics is obeyed for every observer, inside and out \cite{Polhemus:2009vv}.
The black hole also respects Bousso's
covariant entropy bound on the horizon
\cite{Bousso:2002ju}, because the collision entropy never passes through the horizon.  The collision entropy is only produced after the particles have fallen through.
 
The entropy produced in collisions could exceed the total Bekenstein-Hawking radiation after many repetitions (enough to double the black hole's mass).   While this again has no consequences for black hole thermodynamics or entropy bounds, it does raise questions about the number of degrees of freedom available to the black hole.
In an earlier paper \cite{Polhemus:2009vv}
we argue that the excess entropy production inside the horizon
indicates a breakdown of locality inside black holes,
and provides a compelling argument in favor of the conjecture
of ``observer complementarity'' \cite{Susskind:1993if}.
This work was supported by
NSF award
AST-0708607.
We thank the Berkeley Center for Theoretical Physics for their hospitality and useful discussions which inspired this paper.  Thanks also to Tom DeGrand for remediating our understanding of hadron collisions.


\bibliographystyle{JHEP}

\begin{thebibliography}{10}

\bibitem{Wallace:2008zz}
C.~S. Wallace, A.~J.~S. Hamilton, and G.~Polhemus, {\it {Huge entropy
  production inside black holes}},  \href{http://arxiv.org/abs/0801.4415}{{\tt
  arXiv:0801.4415}}.

\bibitem{Polhemus:2009vv}
G.~Polhemus, A.~J.~S. Hamilton, and C.~S. Wallace, {\it {Entropy creation
  inside black holes points to observer complementarity}},  {\em JHEP} {\bf 09}
  (2009) 016, [\href{http://arxiv.org/abs/0903.2290}{{\tt arXiv:0903.2290}}].

\bibitem{Bousso:2002ju}
R.~Bousso, {\it {The holographic principle}},  {\em Rev. Mod. Phys.} {\bf 74}
  (2002) 825--874, [\href{http://arxiv.org/abs/hep-th/0203101}{{\tt
  hep-th/0203101}}].

\bibitem{Flanagan:1999jp}
E.~E. Flanagan, D.~Marolf, and R.~M. Wald, {\it {Proof of Classical Versions of
  the Bousso Entropy Bound and of the Generalized Second Law}},  {\em Phys.
  Rev.} {\bf D62} (2000) 084035,
  [\href{http://arxiv.org/abs/hep-th/9908070}{{\tt hep-th/9908070}}].

\bibitem{Back:2004je}
B.~B. Back {\em et~al.}, {\it {The PHOBOS perspective on discoveries at RHIC}},
   {\em Nucl. Phys.} {\bf A757} (2005) 28--101,
  [\href{http://arxiv.org/abs/nucl-ex/0410022}{{\tt nucl-ex/0410022}}].

\bibitem{Adler:2004zn}
{\bf PHENIX} Collaboration, S.~S. Adler {\em et~al.}, {\it {Systematic studies
  of the centrality and s(NN)**(1/2) dependence of dE(T)/d mu and d N(ch)/d mu
  in heavy ion collisions at mid-rapidity}},  {\em Phys. Rev.} {\bf C71} (2005)
  034908, [\href{http://arxiv.org/abs/nucl-ex/0409015}{{\tt nucl-ex/0409015}}].

\bibitem{Bekenstein:1973ur}
J.~D. Bekenstein, {\it {Black holes and entropy}},  {\em Phys. Rev.} {\bf D7}
  (1973) 2333--2346.

\bibitem{Wald:1999vt}
R.~M. Wald, {\it {The thermodynamics of black holes}},  {\em Living Rev. Rel.}
  {\bf 4} (2001) 6, [\href{http://arxiv.org/abs/gr-qc/9912119}{{\tt
  gr-qc/9912119}}].

\bibitem{Fries:2009wh}
R.~J. Fries, T.~Kunihiro, B.~Muller, A.~Ohnishi, and A.~Schafer, {\it {From 0
  to 5000 in $2\times 10^{-24}$ seconds: Entropy production in relativistic
  heavy-ion collisions}},  {\em Nucl. Phys.} {\bf A830} (2009) 519c--522c,
  [\href{http://arxiv.org/abs/0906.5293}{{\tt arXiv:0906.5293}}].

\bibitem{Gubser:2008pc}
S.~S. Gubser, S.~S. Pufu, and A.~Yarom, {\it {Entropy production in collisions
  of gravitational shock waves and of heavy ions}},  {\em Phys. Rev.} {\bf D78}
  (2008) 066014, [\href{http://arxiv.org/abs/0805.1551}{{\tt
  arXiv:0805.1551}}].

\bibitem{Poisson:1990eh}
E.~Poisson and W.~Israel, {\it {Internal structure of black holes}},  {\em
  Phys. Rev.} {\bf D41} (1990) 1796--1809.

\bibitem{Hamilton:2008zz}
A.~J.~S. Hamilton and P.~P. Avelino, {\it {The physics of the relativistic
  counter-streaming instability that drives mass inflation inside black
  holes}},  \href{http://arxiv.org/abs/0811.1926}{{\tt arXiv:0811.1926}}.

\bibitem{Susskind:1993if}
L.~Susskind, L.~Thorlacius, and J.~Uglum, {\it {The Stretched horizon and black
  hole complementarity}},  {\em Phys. Rev.} {\bf D48} (1993) 3743--3761,
  [\href{http://arxiv.org/abs/hep-th/9306069}{{\tt hep-th/9306069}}].

\end{thebibliography}
\providecommand{\href}[2]{#2}\begingroup\raggedright\endgroup

\end{document}